\documentclass[preprint,prd,tightenlines,superscriptaddress,showkeys]{revtex4}

\usepackage{graphicx} 
\usepackage{dcolumn}  
\usepackage{colordvi}
\usepackage{color}
\usepackage{epstopdf}
\usepackage{subfig}
\usepackage{amssymb}
\usepackage{url}
\graphicspath{{ps}}
\usepackage{hyperref}
\usepackage{tabularx}

\input belle2-sym.tex

\def \Y       {\ensuremath{\Upsilon(4S)}\xspace}

\begin{document}


\def\belletwo {{Belle II}\xspace}
\def\itbelletwo {{\it {Belle II}}\xspace}
\def\phaseiii {{Phase III}\xspace}
\def\itphaseiii {{\it {Phase III}}\xspace}

\newcommand\logten{\ensuremath{\log_{10}\;}}

\vspace*{-4\baselineskip}
\resizebox{!}{3cm}{\includegraphics{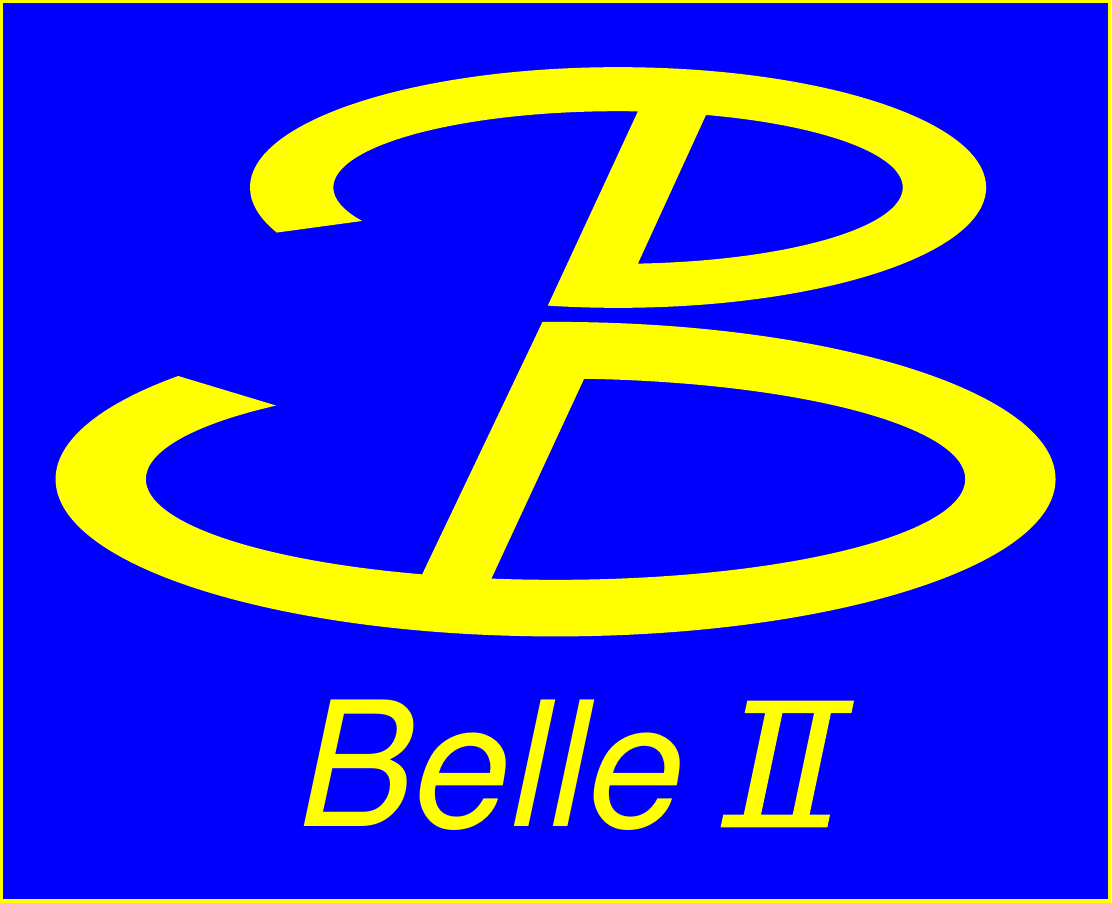}}

\vspace*{-5\baselineskip}
\begin{flushright}
BELLE2-CONF-PROC-2022-006 \\ \today
\end{flushright}
\vspace*{1.8\baselineskip}
{\centering \textit{Presented at the 30th International Symposium on Lepton Photon Interactions at High Energies, hosted by the University of Manchester, 10-14 January 2022.}}

\title { \quad\\[0.5cm] Diversity and inclusion activities in Belle II}

\author{H.M.~Wakeling}
\thanks{Speaker}
\email{hannah.wakeling@physics.mcgill.ca}
\affiliation{Department of Physics, McGill University,\\
3600 rue University, Montr\'eal, Canada}

\author{S.A.~De La Motte}
\email{shanette.delamotte@adelaide.edu.au}
\affiliation{Department of Physics, The University of Adelaide,\\ Adelaide South Australia 5005 Australia}

\author{M.~Barrett}
\email{mattb@post.kek.jp}
\affiliation{Institute of Particle and Nuclear Studies\\
  High Energy Accelerator Research Organization (KEK),\\
  1--1 Oho, Tsukuba, Ibaraki, 305--0801, Japan}

\author{E.~Prencipe}
\email{e.prencipe@fz-juelich.de}
\affiliation{University of Giessen\\  Ludwigstraße 23, 35390 Gießen, Germany}

\collaboration{The Belle II Collaboration}
\noaffiliation

\begin{abstract}

These proceedings accompany the Belle II talk in the Science in Society parallel session delivered during Lepton Photon 2021. In this talk we present updated membership statistics using 10 years of data with a diversity and inclusion lens, and we present Belle II's most recent activities to aid and improve diversity and inclusion. This report has the intention to bring light to the social working environment and population representation within our collaboration and, by extension, within high energy physics.

Belle II is a particle physics collaboration that has over 1000 people
from institutions in 26 countries who work together to achieve its
physics goals. Belle II is committed to fostering an open, diverse, and
inclusive environment; as part of this commitment it created a diversity
office to raise awareness of diversity and inclusion issues, promote an
inclusive atmosphere within the collaboration, provide a safe and
confidential point to contact for collaborators to report any issues,
particularly those related to discrimination and harassment, and ensure
that persons from underrepresented groups are considered for positions
of responsibility within the collaboration. Diversity and inclusion
activities and initiatives at Belle II and analysis of the demographics
of the collaboration will be presented.


\end{abstract}

\maketitle
\section*{Introduction}

Belle II is a high energy physics collaboration located in Tsukuba, Japan. Belle II is the particle detector at the collision point of the asymmetric $e^{+}e^{-}$ SuperKEKB accelerator. At the end of the 2021 Japanese fiscal year (JFYear), Belle II had 1120 active members spread over 4 continents. This paper will introduce the Belle II collaboration and the efforts to research and improve on diversity and inclusion within the collaboration.
Section \ref{sec:CurrentClimate} will present the statistics collected from the collaboration's membership registrations. Results were first presented at ICHEP2020 \cite{wakeling2021diversity}, which included insights into the first Belle II membership survey taken in 2018. The reader is directed to those proceedings for more information on the survey. Belle II has the intention to conduct the next membership survey imminently. 

Section \ref{sec:Actions} will exhibit some of the actions Belle II has been taking, and will take, to promote diversity and inclusion.

\section{Belle II collaboration demographics\label{sec:CurrentClimate}}

This section utilises data from 10 years of Belle II registration data from the Belle II member registration platforms from 2011 - 2021. This data is studied with a diversity and inclusion lens, with a particular focus on gender representation due to the data available for analysis. In 2017, Belle II migrated from a previous system, maintained by the Belle II secretariat, to the Belle II Membership Management System (B2MMS), maintained by Belle II Collaborative Services. Due to this migration, results for 2017 may appear anomalous and should not be taken as indicative of real trends.

Within the B2MMS, it is required that a member submit their name, title, gender, email, institution and membership category (i.e. Ph.D. student, faculty, etc.) to register as part of the Belle II collaboration. As of January 2022, gender is no longer required to be defined as either `Male', `Female' or `Other'; an option for the field to be left blank has been added to allow members not to disclose their gender. Here we acknowledge that this does not allow non-cisgender members to specify their gender if they wish to, and possible solutions to this will be discussed at future diversity and inclusion meetings. This form of data taking biases data and can make it hard for non-cisgender members to feel welcomed into our community before they have even officially joined the collaboration. Currently no member has declared themselves as `Other' in the B2MMS. If there are indeed non-binary Belle II members, this lack of declaration indicates difficulties in gender affirmation within the workplace or due to legal reasons.

Additionally, in the following figures there is an `unspecified' category; this is a relic from the 2017 changeover of systems where the gender field was empty in the pre-B2MMS system. In the following histograms it is necessary to combine `other’ and `unspecified’ into one category as underrepresented persons are identifiable due to low statistics. We as a collaboration will maintain and pursue true anonymity of our membership data. 
\begin{figure}[h]
    \centering
    \begin{minipage}{.45\textwidth}
        \captionsetup{width=.95\linewidth}
        \includegraphics[width=\linewidth]{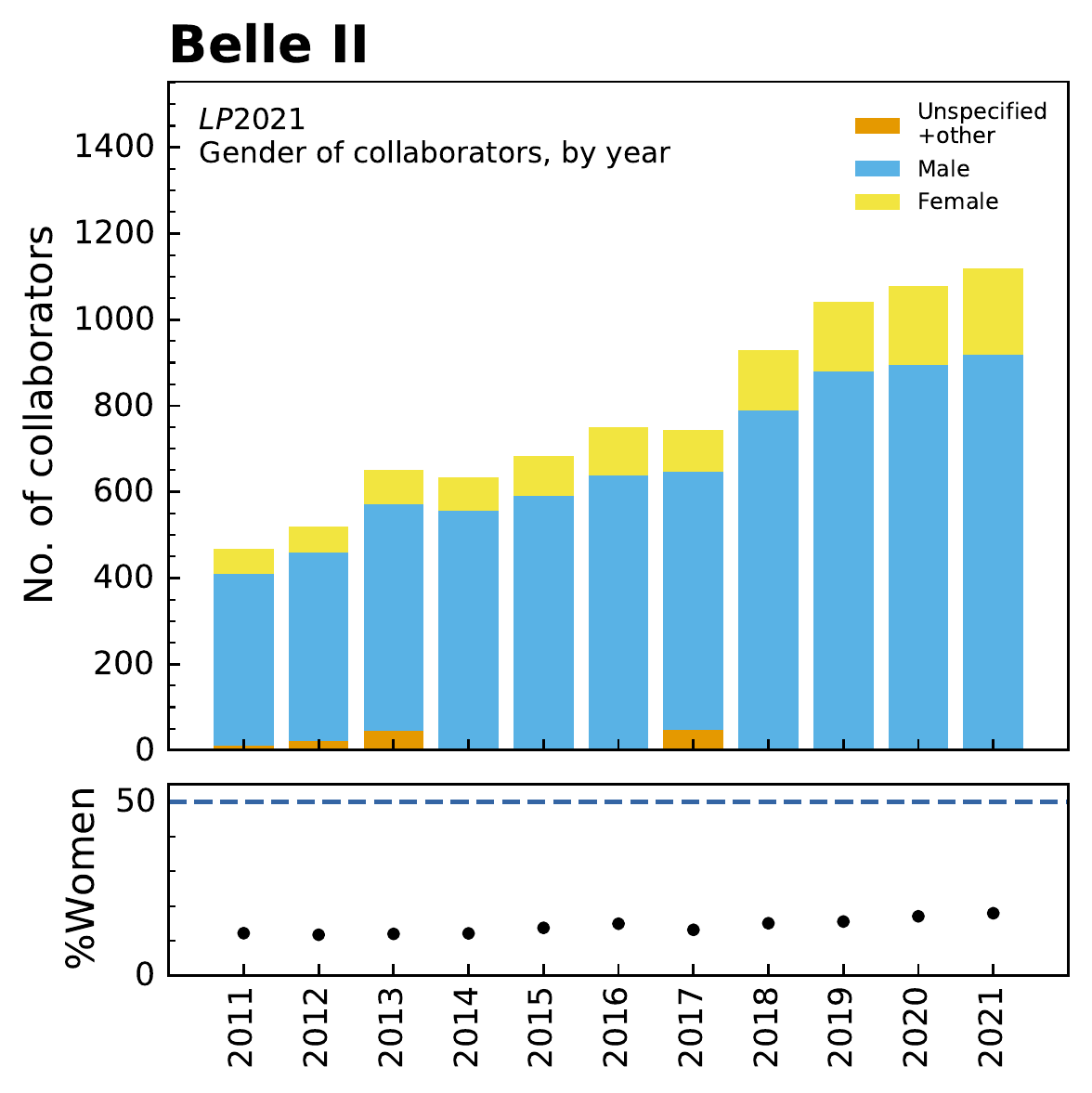}
        \vspace*{1em}
        \caption{The number and gender of Belle II Collaborators, from years 2011 to 2021. Data taken from Belle II membership declaration, where `other' refers to non-binary gender identity and `unspecified' refers to undeclared gender information. \newline \newline}
        \label{fig:by_year}
    \end{minipage}%
    \begin{minipage}{.45\textwidth}
        \captionsetup{width=.95\linewidth}
        \includegraphics[width=\linewidth]{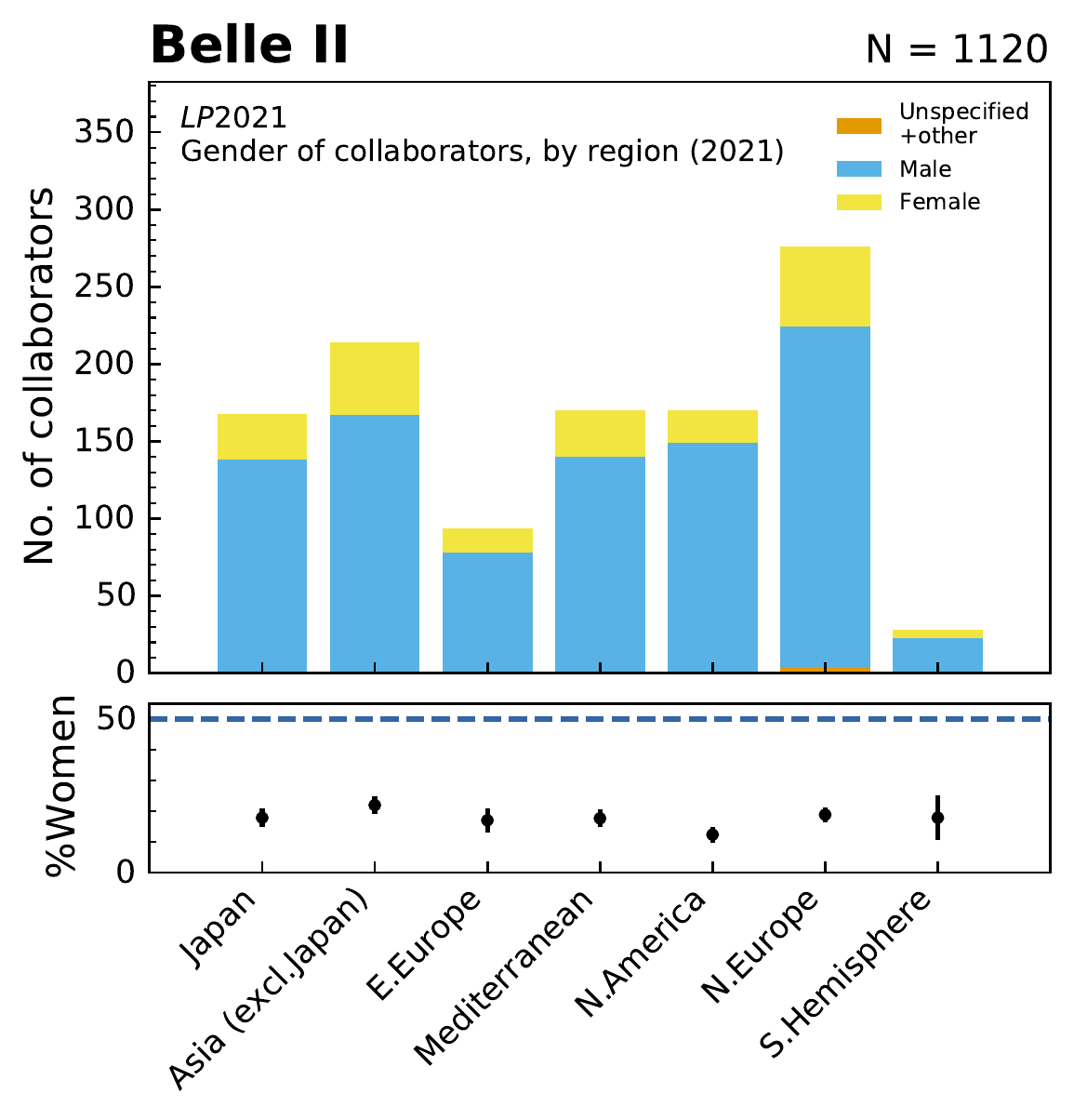}
        \caption{The number of collaborators of the Belle II Collaboration by region and gender.}
        \label{fig:by_region}
    \end{minipage}
\end{figure}

Between the years 2011 to 2021 the number of total members of Belle II has approximately tripled. In 10 years, the proportion of female-identifying members has approximately quadrupled from 57 members in 2011 to 201 members (Figure \ref{fig:by_year}). However, the percentage of female-identifying members within the collaboration is still only increasing at an average of 0.57\% each year (calculated from the overall percentage increase of 5.7\%, from 12.2\% to 17.9\%). In comparison to the ICHEP2020 reported rate of 0.4\%, Belle II is moving in the right direction with an increase in rate of the number of female-identifying members. However this is insufficient as, at this rate, Belle II would not reach gender parity within the anticipated lifetime of the Belle II experiment. 

It is beneficial to consider the effects of the age of the Belle II collaboration. As the collaboration is relatively young, the growing needs for personpower to do operational activities and analyses rapidly accelerated. Over the first decade of the experiment, the average number of students and postgraduates increased, and the average age of the collaboration decreased. Thus, this observed change in gender demographics may have changed at a different rate than that of a long-established experiment.

We next compare the gender of collaborators by the region they work in, based on their registered institution. When compared to similar graphs produced by the ATLAS Collaboration \cite{ATL-GEN-PUB-2016-001} we see that we have a much larger proportion of collaborators from Asian institutions than ATLAS.
Japan alone contributes 15.00\% of the collaboration, so the category is separated from the ‘Asia’ region for clearer presentation. One particularly interesting observation can be taken from Figure \ref{fig:by_region}, through the comparison of regions with approximately equal numbers of members and therefore similar representation within the collaboration. Comparing Japan, which represents 15.00\% of the collaboration, to Northern America which represents 15.18\% of the collaboration, we can see that Japan has a higher proportion of women than Northern America, almost 17.86\% compared to 12.35\% respectively.
\begin{figure}[h]
    \centering
    \begin{minipage}{.5\textwidth}
        \captionsetup{width=.9\linewidth}
        \includegraphics[width=\linewidth]{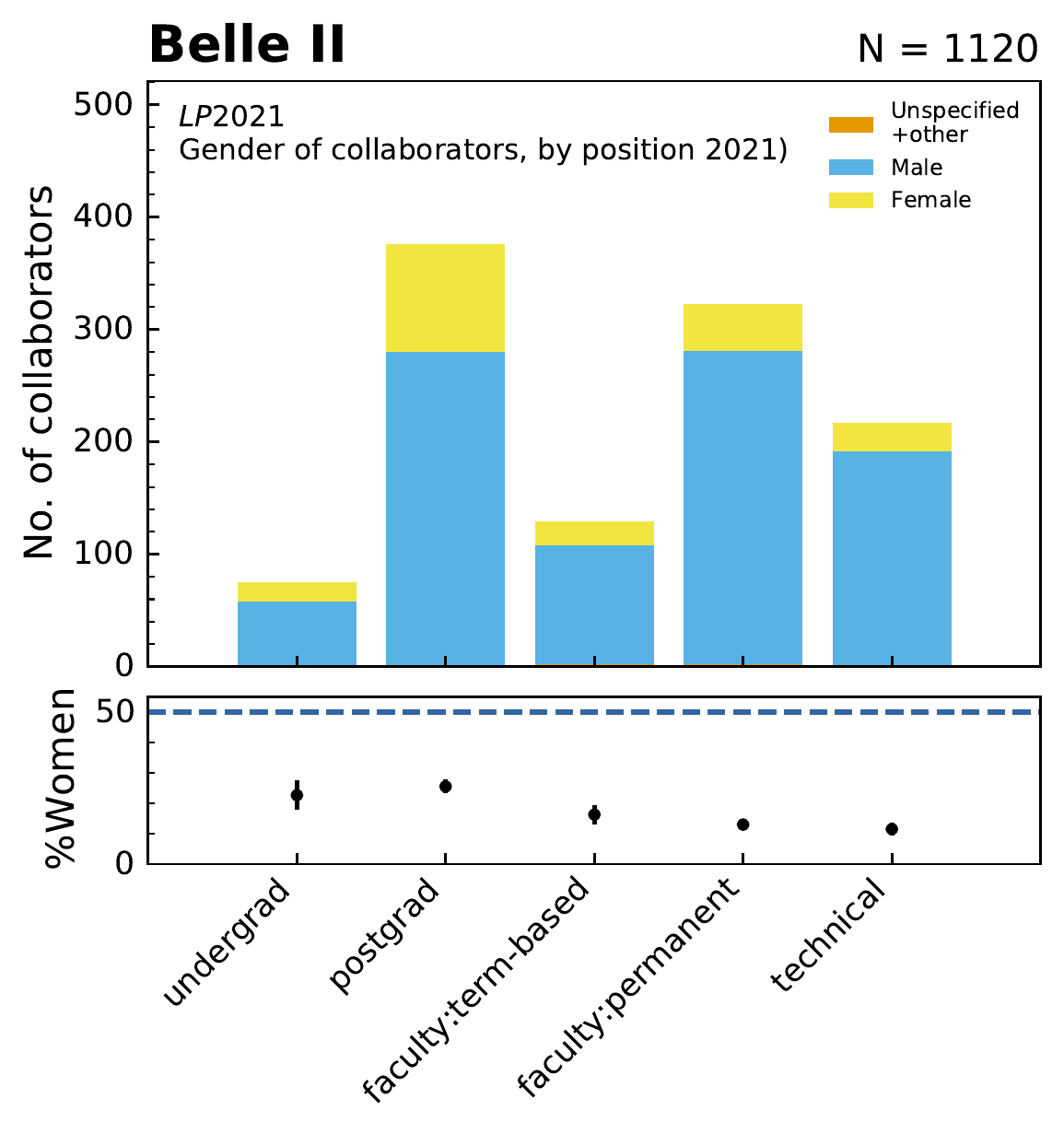}
        \caption{The number of collaborators of the Belle II Collaboration by position and gender.}
        \label{fig:by_pos}
    \end{minipage}%
    \begin{minipage}{.5\textwidth}
        \captionsetup{width=.9\linewidth}
        \includegraphics[width=\linewidth]{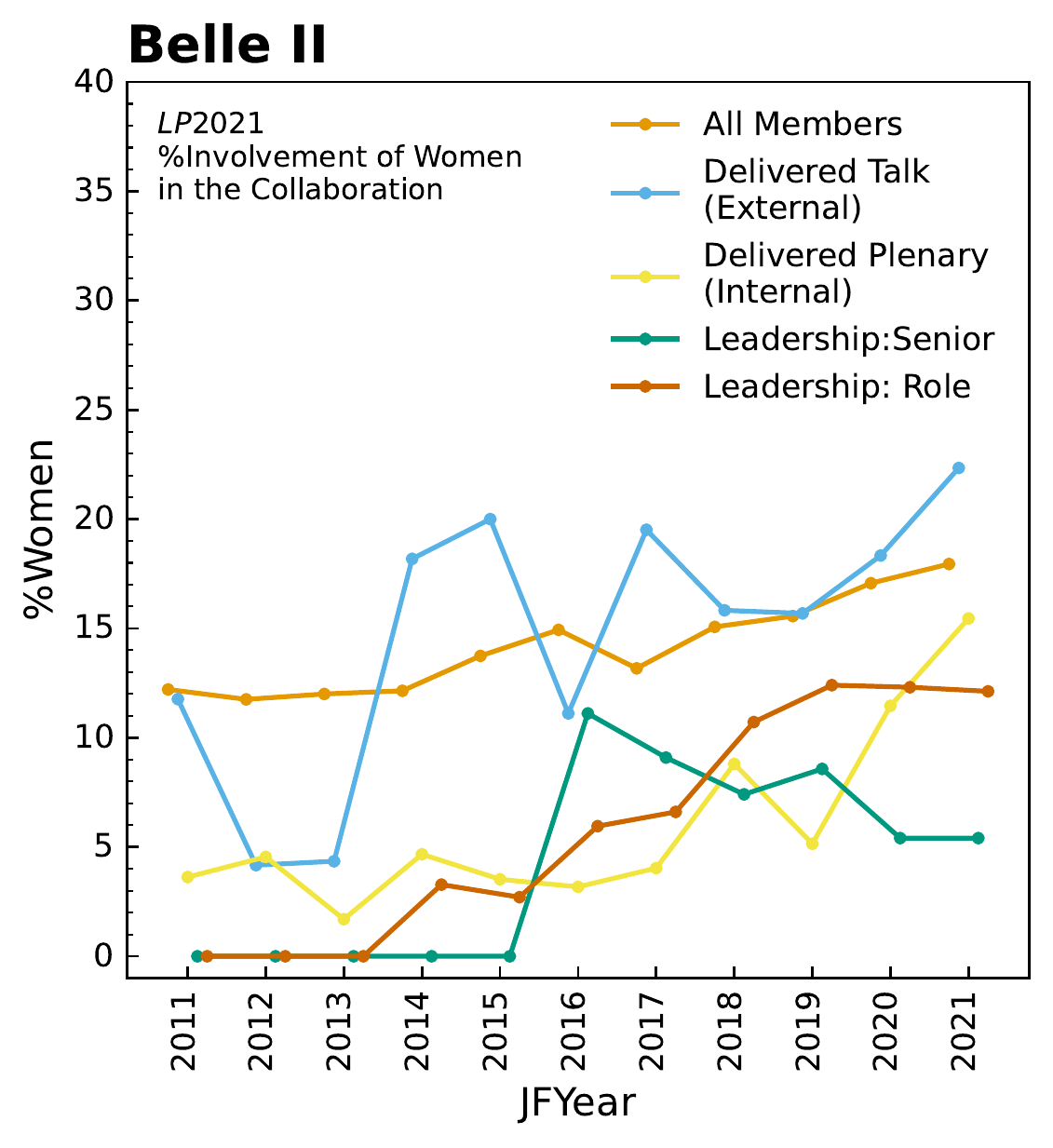}
        \caption{The percentage of people in the specified roles or recognised involvement who identify as women.}
        \label{fig:ratio_scatter}
    \end{minipage}
\end{figure}
Within Belle II, the representation of women as a percentage drops as their careers progress from postgraduate to permanent faculty, shown in Figure \ref{fig:by_pos}. The under-representation of women increases as their career progresses from postgraduate to permanent faculty. This effect is likely influenced by the age of members who are in their later career stages. The increase in percentage of people who identify as underrepresented minorities in science has been gradual over the last few decades. Though this effect may exist at Belle II, we are not entirely exempt from the coined (and passively termed) `leaky pipeline’ effect, as there is a drop in gender representation at the permanent faculty level.


Figure \ref{fig:ratio_scatter} shows the percentage of women in recognized involvement in the collaboration by year. The figure shows an upwards trend over the last 10 years of data collection, but leaves a lot of room for improvement. Internal talks are often given by those in leadership roles such as group chairs. These positions have a lower percentage of female representation due to many factors, including the fact that leadership roles tend to be held by more senior members, therefore the fact that there are even fewer women doing internal talks is not surprising.

\section{Belle II diversity and inclusion actions \label{sec:Actions}}

Having a large, culturally diverse collaboration has led to Belle II establishing a code of conduct. Belle II included the first iteration of the code of conduct in the Belle II bylaws in October 2017 and then updated them in 2018. More details can be found in Reference \cite{wakeling2021diversity} and on the Belle II diversity webpage \cite{website}.

Belle II uses its active social media platforms to raise awareness of diversity and inclusion events such as International Women’s Day, LGBTSTEM Day, and Colour Blind Awareness Day. The posts are always published in both Japanese and English. 
In 2019, Belle II became an official supporter of LGBTSTEM Day, through unanimous endorsement of our Institutional Board. Belle II changes its profile picture for the day to a logo with an LBGTQ+ rainbow flag background, and also makes a skin available for profile pictures on Facebook. We encourage other research institutions to support each LGBTSTEM day.

Belle II has been working on a multitude of other new projects to improve equity and inclusion within the collaboration. For example, Belle II has been improving certain language used in computing and physics. In general, the collaboration wishes to avoid anything that might cause distress or feelings of exclusion to our collaborators. In particular, words with severe racial overtones are either already phased out or in the process of being phased out. Additionally, the Belle II eCafe project has just begun; an initiative in which junior researchers have a biweekly space to chat to peers and discuss issues with the diversity officers. Belle II has also implemented colour blind friendly screens in the experiment control room; analysts and users are encouraged to use colour blind friendly colours in plots and all plots in this conference note have adhered to these schemes. Finally, a men-only bathroom adjacent to the Belle II control room has been renovated into a gender neutral, accessible bathroom. Previously the men's bathroom was the only available bathroom on the floor and women would have to change floors for bathroom access; there was no designated access to non-cisgender bathrooms and the women's bathroom was not accessibility friendly.

Belle II will continue to work within the collaboration, to improve accessibility, to foster an inclusive work environment and to provide services that encourage diversity, equity and inclusion. We wish to hold ourselves accountable, and aims to provide regular updates on this work and on member statistics. Previous reports are available: EPS 2021 \cite{DeLamotte2022diversity}, and ICHEP 2020 \cite{wakeling2021diversity}. We thank the Belle II Secretariat, Belle II Collaborative Services, and the Belle II Speakers Committee for their invaluable assistance and for maintaining Belle II membership and conference statistics.

\bibliographystyle{unsrt}
\bibliography{references}
\end{document}